\documentclass[showpacs,showkeywords,pof,reprint,preprintnumbers,amsmath,amssymb]{revtex4-1}
\pdfoutput=1
\usepackage[pdftex]{graphicx}
\usepackage{mathrsfs,mathcomp}
\usepackage{dcolumn}
\usepackage{bm}

\begin{document}

\title{Building water bridges in air: Electrohydrodynamics of the Floating Water Bridge}

\author{\'Alvaro G. Mar\'in}
\author{Detlef Lohse}
\affiliation{Physics of Fluids, University of Twente}

\date{\today}


\begin{abstract}\label{abstract}

The interaction of electrical fields and liquids can lead to phenomena that defies intuition. Some famous examples can be found in Electrohydrodynamics as Taylor cones, whipping jets or non-coalescing drops. A less famous example is the Floating Water Bridge: a slender thread of water held between two glass beakers in which a high voltage difference is applied. Surprisingly, the water bridge defies gravity even when the beakers are separated at distances up to 2 cm. In the presentation, experimental measurements and simple models are proposed and discussed for the stability of the bridge and the source of the flow, revealing an important role of polarization forces on the stability of the water bridge. On the other hand, the observed flow can only be explained due to the non negligible free charge present in the surface. In this sense, the Floating Water Bridge can be considered as an extreme case of a leaky dielectric liquid (J. R. Melcher and G. I. Taylor, Annu. Rev. Fluid Mech., 1:111, 1969).

\end{abstract}

\maketitle
\section{Introduction} \label{intro}

Electromagnetic fields can only be visualized through its interaction with matter. Such a manifestation of electromagnetic forces often defies intuition. Some paradigmatic examples can be found in Electrohydrodynamics and have been well studied in the literature: Taylor cones \cite{Taylor:1964}\cite{Mora:2007}\cite{Basaran:2008}, electrohydrodynamic driven whipping jets\cite{Taylor:1969}\cite{HohmanBrenner:2001}\cite{Riboux2010} or anti-coalescent drops\cite{BirdStone:2009}. But a few phenomena remain still without convincing explanations. The purpose of the present work is to analyze and provide reasonable explanations to one of these phenomena: The so-called ``floating water bridge'' is formed between two glass beakers full of purified water when an electrical high voltage difference is applied. Surprisingly, the water bridge defies gravity even when the beakers are separated by distances up to $2 cm$. The experiment is easy enough to reproduce, needing only standard demineralized water and a high voltage power supply (able to give 20kV at low amperage). Due to its relative simplicity and its spectacular features, the phenomenon has become popular in science fairs, videos in the web, forums and some recent publications, specially those by Fuchs et al. \cite{Fuchs:2007}\cite{Fuchs:2008}\cite{Fuchs:2009}\cite{Fuchs:2010}, in which different experiments employing thermal imaging, LDA, Schlieren visualization and neutron scattering were performed on the floating water bridge.
The first reference of a controlled experiment dates back to 1893 \cite{Armstrong:1893}, when the English engineer Lord Armstrong presented a modified version of the floating water bridge in a public presentation, among some other experiments involving high voltages and fluids.
A similar phenomenon has been intensively studied in the literature, the ``dielectric liquid bridge'': A liquid bridge of oil, surrounded by a second immiscible and insulating liquid, is sustained vertically between two parallel plates, in the absence of an electrical voltage, the liquid bridge would break into droplets for values of the aspect ratio (defined as the bridge length to diameter ratio) higher than $\pi$, due to the minimization of the surface in the presence of capillary sinusoidal instabilities \cite{Rayleigh} \cite{Plateau}. When the electrical field is set, the liquid bridge is found to be stable for aspect ratios up to 6 \cite{Gonzalez1989}\cite{Ramos:1994}. Several theoretical and numerical papers \cite{Saville:2002} accounted for the experimental observations, based on the theories developed decades before by G. I. Taylor \& J. R. Melcher\cite{Taylor:1969}.

The aim of this paper is to connect both phenomena and propose an explanation to the floating water bridge in the framework of Electrohydrodynamics. The striking stability of the bridge has two different features: on the one hand the water bridge seems to defy gravity showing an almost horizontal profile, and on the other hand, it resists the break-up into droplets due to capillary forces until extremely large aspect ratios. Both effects are connected and different experiments will be carried out to test them. The first set of experiments are performed in the ``beakers configuration'' \cite{Fuchs:2007} in order to characterize the floating water bridge experiment in terms of Electrohydrodynamic dimensionless numbers, and the second will be done in the so called ``axisymmetric configuration'', in which both aspects of the stability of the water bridge will be analyzed: the stability against capillary forces and the stability against gravity.

\begin{figure}[h!]
\includegraphics[width=3.4 in]{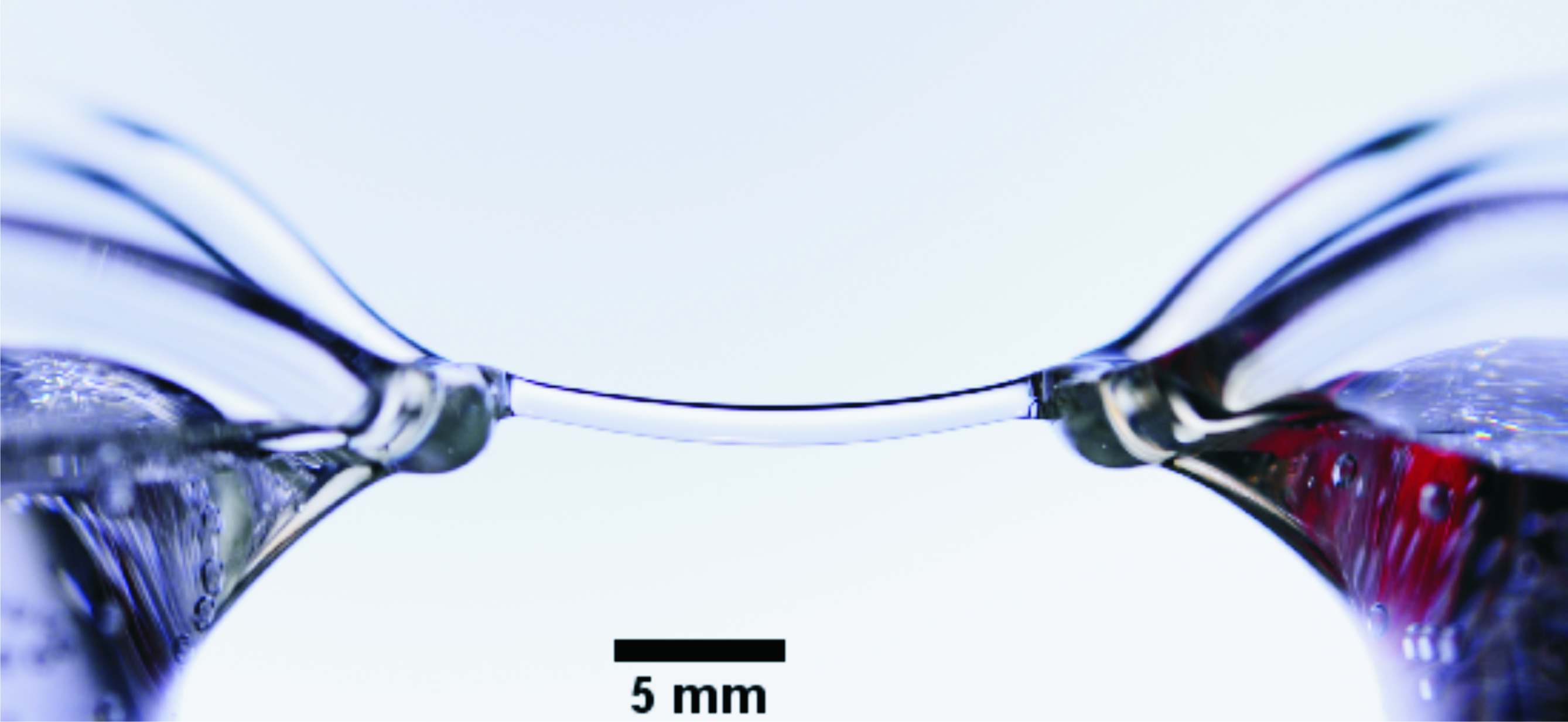}
\caption{ Side view of the Floating water bridge \label{fig:cover}}
\end{figure}

\section{Experiments on the Beakers configuration}\label{section:beakers}

\begin{figure}
\includegraphics[width=3.4 in]{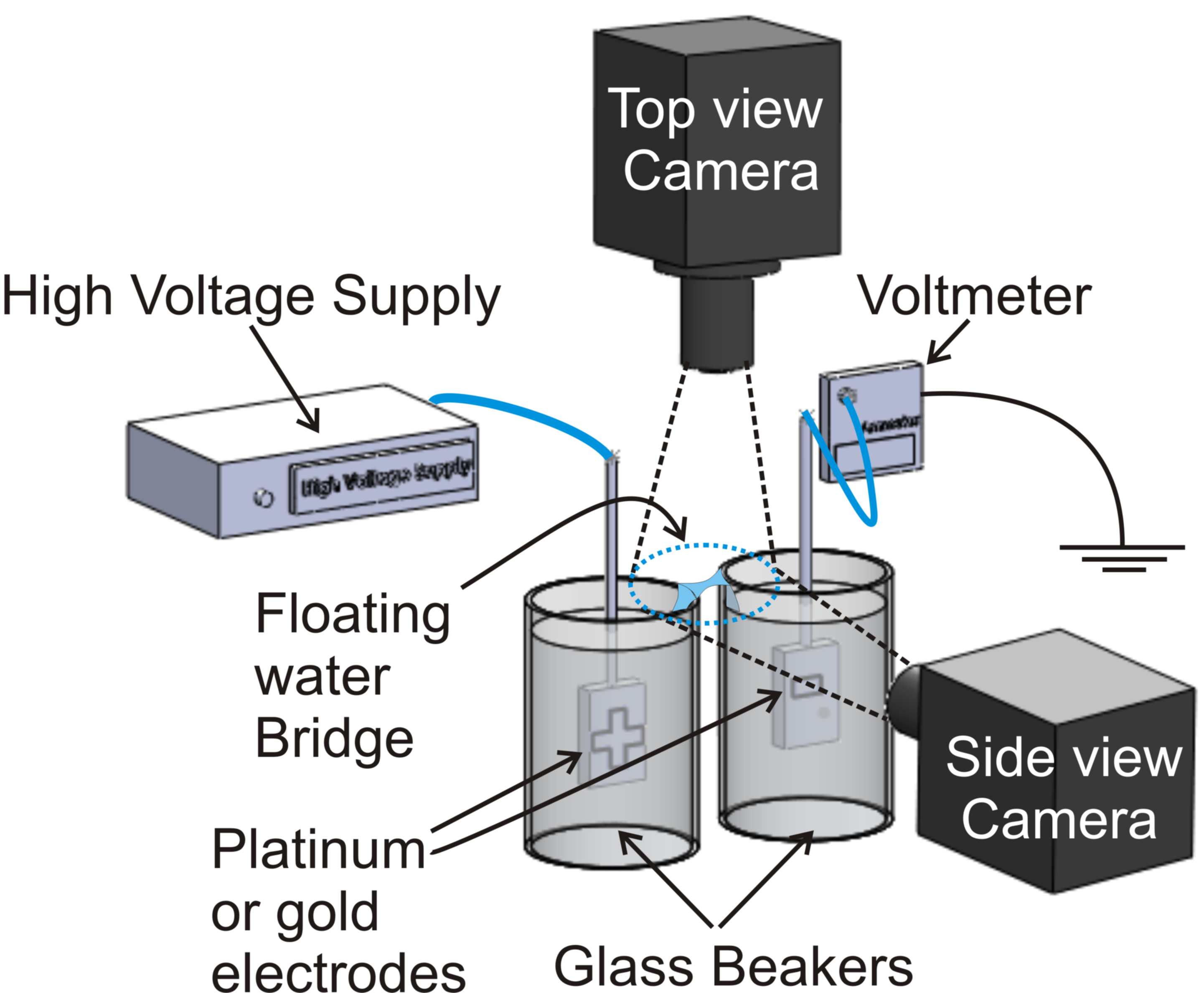}
\caption{Set up for the Beakers configuration\label{fig:setup-beakers}}
\end{figure}

The basic phenomenon is shown in what we will call the ``beakers configuration'' (figure \ref{fig:setup-beakers}): Two glass beakers are filled with liquid, with a flat or cylindrical electrode immersed in each one, preferably covered with platinum or gold, to reduce reactions or corrosion. To avoid the formation of sparks or discharges in air, which can increase the contamination of the liquid, the two beakers are approached until their borders are in contact and filled with the liquid close to the beakers' border. When the electrical voltage is increased up to $10 kV$, water will climb the remaining distance until reaching the beaker border due to the so-called ``Pellat effect'' \cite{Pellat}\cite{jones:2005}\cite{Wang2005}; a film of liquid connects the beakers and its thickness increases as the electrical voltage difference between them is increased. The film thickness increases from tenths of a millimeter to some millimeters when the electrical voltage is high enough ($15-20kV$). Once the beakers are separated, the bridge will remain stable for almost a complete hour, depending on several factors but all related with the liquid purity. For a fixed voltage we can increase the bridge length by separating the beakers; the bridge then becomes slenderer and thinner, until a critical distance is achieved where the bridge pinches off due to capillary instabilities. Similarly, for a fixed bridge length or beakers separation we can reduce the voltage until the bridge bends and collapses by its own weight.

\begin{figure}
\includegraphics[width=3.4 in]{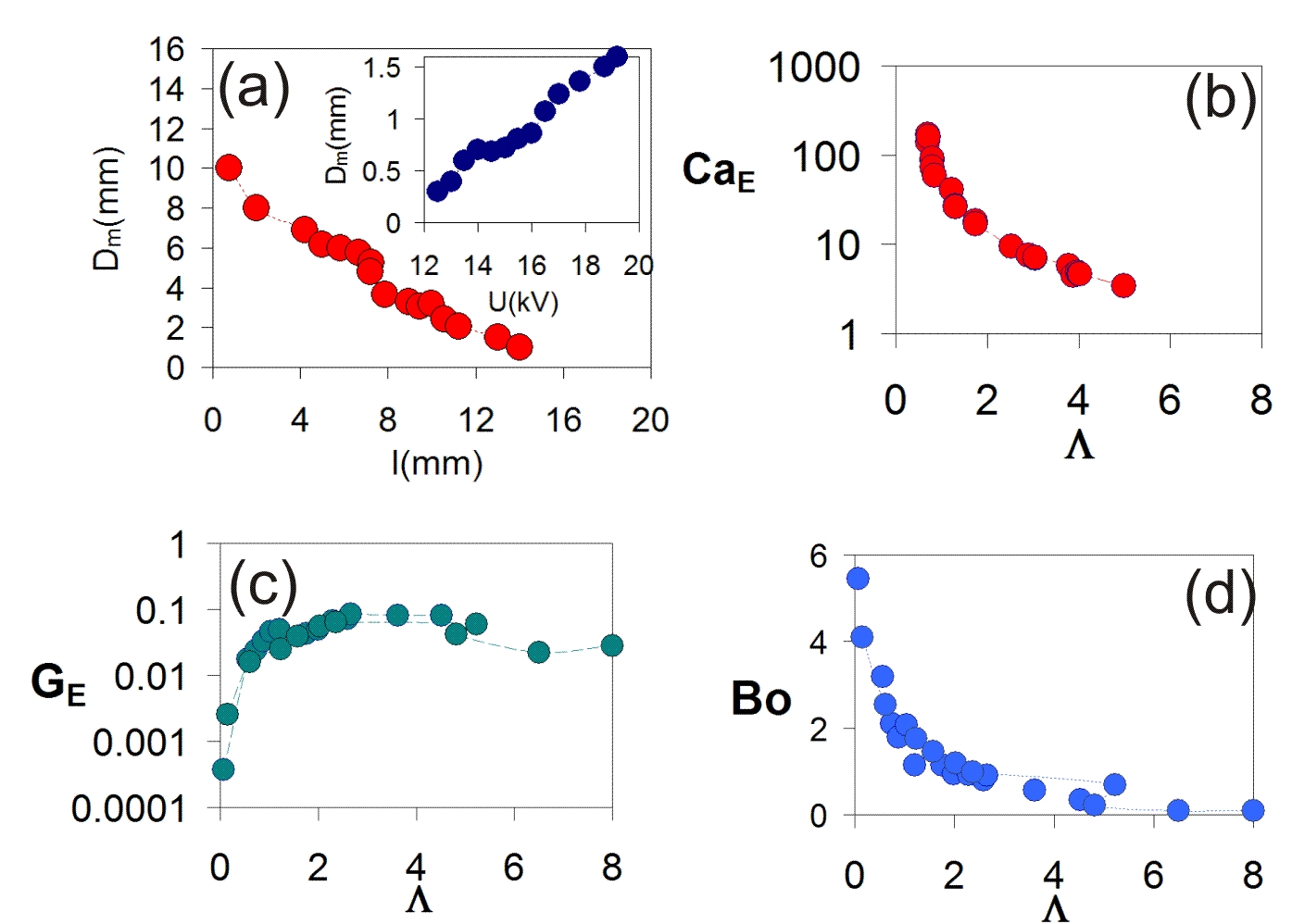}
\caption{Measurements in the beakers configuration: a)raw data for $D_m$ as a function of the beaker separation $l$, and as a function of the applied voltage $U$ (insert). b) Electrocapillary number $Ca$ vs the aspect ratio $\Lambda$ during the separation of the beakers. c) Electrogravitational number $G_E$ vs the aspect ratio $\Lambda$. d) Bond number $Bo$ vs the aspect ratio $\Lambda$ during the separation of the beakers.\label{fig:beakersgraph}}
\end{figure}

In order to analyze the relative importance of the forces being applied to the bridge, the experiments are performed visualizing the shape of the bridge from top and side views at different beaker separations in order to characterize its geometrical characteristics. The governing parameters of the problem can be classified in the following way. The relevant properties of the liquid: $\gamma$ is the surface tension, $\epsilon_r$ is the relative dielectric permittivity, $\rho$ is its density, $\mu$ the viscosity. The geometrical characteristics of the liquid bridge: $l$ is the length of the bridge, defined as the distance between the beakers' tips; $D_m$ is defined as $D_m=\sqrt{{d}_s {d}_t}$, where ${d}_s$ and ${d}_t$ are defined as the bridge diameter obtained from side and top view respectively, arithmetically averaged along their length; $V$ is the bridge volume, defined as $V=\pi{d}_t {d}_s l/4$; The bridge aspect ratio is defined as $\Lambda=l/D_m$. The fields applied to the bridge: $g$ is the gravitational acceleration and $E$ is the electrical field across the bridge, defined as $E_t=U/l$ where $U$ is the nominal voltage difference applied from the voltage supply.
In the present experiment, triply demineralized water was used as a liquid. The control experimental parameters are the nominal voltage $U$ and the bridge length $l$ (controlling the beakers separation). Through top and side views the geometrical characteristics of the bridge: $d_s$ and $d_t$ (and thus $D_m$ and $V$), they can be considered as a response of the system and are accurately measured using a custom-made image processing MATLAB code from high resolution images. To illustrate the dimensions in which the experiments are being run, we include in figure \ref{fig:beakersgraph}(a)raw data for the evolution of $D_m$ for separating beakers (increasing $l$) and increasing $U$. With these parameters, the following dimensionless groups can be defined:

\begin{align}
Ca_E&=\frac{\epsilon_o\epsilon_r E_t^2D_{m}}{\gamma}, & Bo&=\frac{\rho g V^{2/3}}{\gamma}, &
\label{eq:numbers1}
\end{align}

\begin{align}
G_E&=\frac{\rho g V^{1/3}}{\epsilon_o\epsilon_r E^2}=\left(\frac{4}{\pi \Lambda}\right)^{1/3}\frac{Bo}{Ca_E}.
\label{eq:numbers2}
\end{align}

$Ca_E$ is the Electrocapillary number, defined as the ratio between electrical and capillary forces; $Bo$ is the Bond number, accounting for the balance between gravity and capillary forces \ref{eq:numbers1}. Finally, the ratio of gravitational and electrical forces can be expressed in terms of the Electrogravitational number $G_E$ \ref{eq:numbers2}, which can be expressed in terms of $Ca$, $Bo$ and $\Lambda$. One of the main issues in this configuration is the way to estimate the electric field at the bridge surface. As a rough approximation, $E_t$ is taken as the ratio of the applied voltage $U$ over the beaker separation $l$. For small beakers separation, the real electrical field at the bridge surface is expected to be smaller than $E_t$. Therefore such an approximation will clearly overestimate the values the Electrocapillary number and underestimate the Electrogravitational number for small beakers separation, but nonetheless will serve for the purpose of this section.

\begin{figure}
\centering
\includegraphics[width=3.4 in]{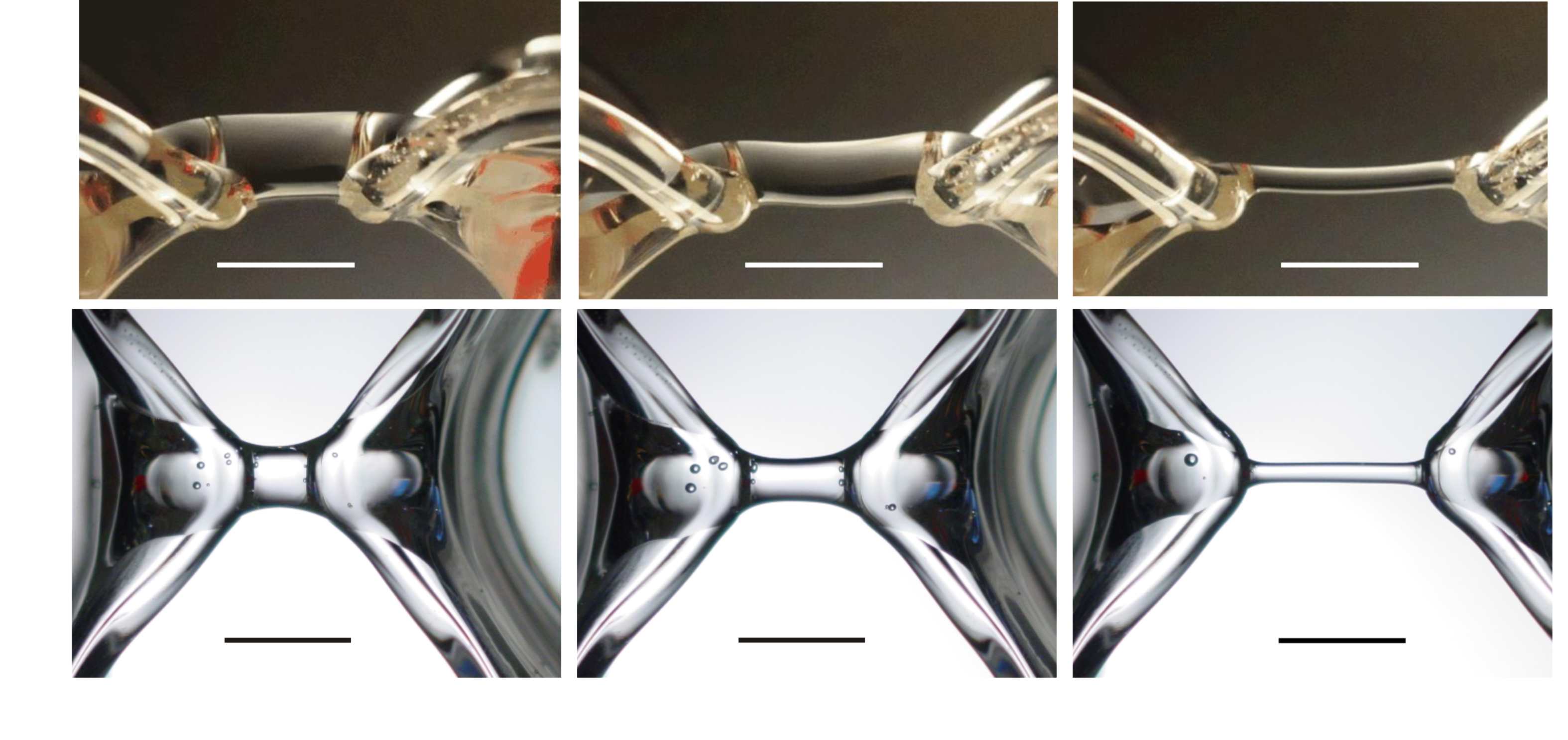}
\caption{First row of images: Different side views during separation of the beakers. Second row: Different top views during separation of the beakers. Nominal voltage 15kV. The scale bars have a length of $10 mm$; in the right images, the bridge reaches an aspect ratio $\Lambda \approx 10$. \label{fig:beakersmontage}}
\end{figure}

In figure \ref{fig:beakersgraph}(a) some raw data of $D_m$ during the separation of two water beakers at 15kV is plotted, while in the insert of the same figure we show raw data of $D_m$ for decreasing voltages at a fixed bridge length of 14mm (similar as that shown in figure \ref{fig:cover}). The data points with the longest length and the lowest voltage approximately correspond to the breakup points respectively on each graph. Although the electrical current has not been shown for clearness in this last experiment, we should mention that it decreases linearly for decreasing voltages from values of order 0.8mA to nearly zero when the bridge collapses. Since the current passing through the bridge is clearly non-negligible, the bridge suffers Joule heating. However, the experiments shown in figure \ref{fig:beakersgraph} were performed normally in short sessions of maximum 5 min, after which the water was replaced. The measured temperature at the end of each short session was close to $45\tccentigrade$ ($113^{\circ}$F), this would imply an error in the surface tension of around $4\%$. For a fixed voltage $U$ and increasing beaker distances $l$, the values of these dimensionless numbers are plotted in figures \ref{fig:beakergraph}(b) to (c). As can be seen in figure \ref{fig:beakersgraph}(b), the Electrocapillary number $Ca_E$ decreases until values still above unity in the breakup point, in figure \ref{fig:beakersgraph}(c) the Electrogravitational number $G_E$ increases, but remains well below unity during the whole process, and finally in figure \ref{fig:beakersgraph}(d) the Bond number $Bo$ and the Electrocapillary number $Ca_E$ decrease until the bridge collapses with Bond numbers well below unity. This shows us that the initial stage of the bridge (left column in figure \ref{fig:beakersmontage}) is dominated by Electrical forces, with negligible effect of the capillary forces and gravity. In contrast, in the late stages when the bridge becomes thinner (right column in figure \ref{fig:beakersmontage}) there seems to be a delicate balance of electrical and capillary forces, and therefore the Electrocapillary number is the most relevant number in the most elongated bridges.


From this point of view, this last stage shares similar characteristics with the classical work on dielectric liquid bridges under electrical fields \cite{Gonzalez1989}\cite{Saville:2002}. The shown results are interesting to identify the main forces, but in order to proceed with a more detailed analysis one must be able to determine more precisely the electric field close to the bridge interface. For this reason, the ``Axisymmetric configuration'' is employed in the following, in which the struggle of the electrical forces against capillarity and gravity are studied.

\section{Axisymmetric configuration: stability against capillary forces}

The stabilization of dielectric liquid cylinders under the action of a longitudinal electrical fields have been studied theoretically-numerically \cite{Nayyar1960} \cite{Saville:2002} and experimentally \cite{Gonzalez1989} \cite{Ramos:1994}. The underlying physics are based on induced polarization forces on dielectrics \cite{Melcher:book} \cite{CastellanosEHD}: in a pure dielectric liquid cylinder under an electrical field applied parallel to its interface, any sinusoidal perturbation developed over its surface would create polarization charges of opposite signs in different slopes that will tend to stabilize the surface. On the other hand, for electrically conducting liquids, the charge in semi-equilibrium over the surface makes the picture much more complicated and the equilibrium is no longer guaranteed. Since liquids in nature are neither pure conductors nor pure dielectrics, Taylor and Melcher developed the so called ``leaky dielectric model''\cite{MelcherTaylor:1969} \cite{Saville:1997} in which a dielectric liquid is assumed to have some free charge that only manifests at its surface. Such a model was applied to a leaky dielectric liquid cylinder under a longitudinal electrical field by Saville \cite{Saville:2002}, concluding that surface charge transport would lead to a charge redistribution and a consequent instability of the liquid cylinder. A similar but not comparable result\cite{Saville:1971} was found on infinite jets: unstable but oscillatory modes were found as the charge relaxation effects, i.e. the ability of the free charge to find its equilibrium on the surface, were increased.


\begin{figure}
\centering
\includegraphics[width=3.4 in]{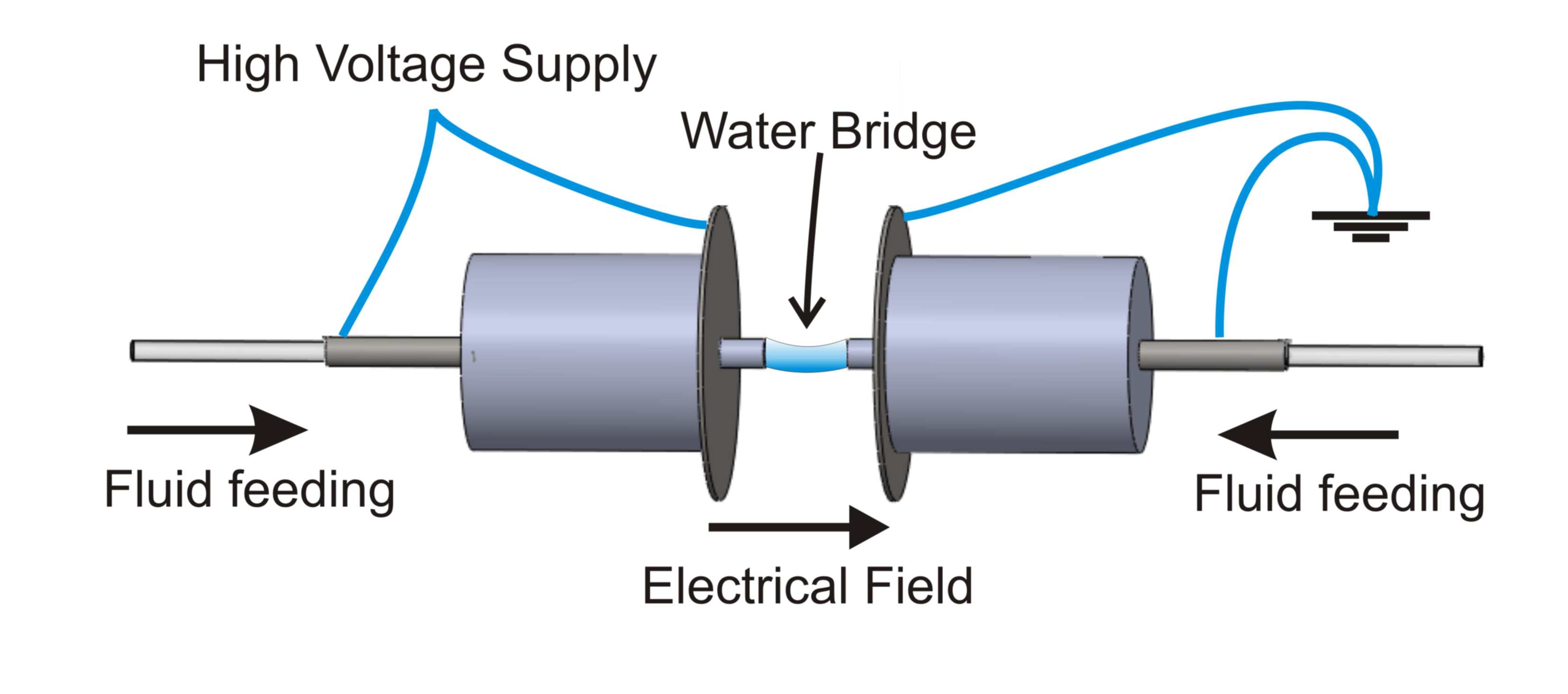}
\caption{Set up for the Axisymmetric configuration\label{fig:setup-axys}}
\end{figure}

\begin{table*}[tp]\footnotesize
\caption{Properties of the employed liquids}\label{table1}\centering
\begin{tabular}[c]{|m{4.5cm}|m{1.5cm}|m{1.5cm}|m{1.6cm}|m{1.5cm}|m{2.1 cm}|}
	\hline
Liquid &  Density $\rho\;(kg/m^3)$  & Surface Tension $\gamma\;(mN/m)$
   & Dynamic Viscosity  $\mu\:(mPa\cdot s)$ & Dielectric Constant  $\epsilon_r$ & Electrical Conductivity $K\;(\mu S/cm)$ \\
	\hline

Glycerine                       &   1250    &  64	& 1050      &  41.6    &   0.015   \\
75\%w Glycerine- 25\%w water    &   1182    &  66	&	31      & 54.7     &   0.213   \\
50\%w Glycerine- 50\%w water    &   1115	&  68	&    5      & 65.6     &   0.560   \\
Ultrapure millipore water       &   980     &  72   &   1       &   81     &   0.520   \\

	\hline
\end{tabular}
\end{table*}

Therefore, if the induced polarization forces are responsible for the stability of the liquid bridge, less electrical voltage $U$ will be needed to maintain a liquid bridge as the dielectric permittivity $\epsilon_r$ of the liquid increases. In order to test such an effect, an axisymmetric set up was built, basically consisting of two metallic parallel plates through which two plastic nozzles protruded. The liquid is dispensed through the plastic nozzles, of $2 mm$ inner diameter and $3 mm$ outer diameter. Each liquid line is connected to voltage through either a metallic tube or a small platinum electrode, in both cases allocated close to the end of the nozzles. Both nozzles are precisely placed in front of each other by two 3D micropositioners (New Focus), which permit us to change the bridge length $l$ with high precision. The electrical voltage $U$ is set by a high voltage power supply Glassman HV, model FC30R4 120W (0-30kV, 0-4mA), and is operated for a fixed and known voltage. The current is measured during the whole process, making sure that the voltage drop along the bridge is not lower than a $90\%$ of the applied one. The axisymmetric set up has several advantages: first of all, the external electrical field is controlled independently and can be assumed to be axisymmetric and homogeneous. Secondly, the volume in the liquid bridge is now precisely controlled by two syringe pumps (PHD 2000 infusion syringe pump, Harvard Apparatus). And finally, the total electrical resistance can also be controlled by modifying the distance of the electrodes to the exit of the nozzles.

For probing the effect of the dielectric permittivity, three different liquids mixtures of glycerin and pure water were used of dielectric constants of 41.6 (pure glycerine), 54.7 (25\% water-75\% glycerin) and 65.6 (50\% water - 50\% glycerin). The physical properties can be found in table \ref{table1}. In the axysimmetric configuration, the control parameters are the length of the bridge $l$, the bridge volume $V$, the nominal voltage $U$ and also the dielectric permittivity of the liquid $\epsilon_r$. The experimental protocol was the following: for a given electrical voltage difference $U$, both nozzles were slowly separated (increasing $l$) until the liquid bridge breaks. In this moment, $Ca_E$ and the critical aspect ratio $\Lambda_c$ were calculated. The process was filmed with a CCD digital camera (Lumenera corp model LW135M), being triggered and synchronized with the measurements of the electrical current through the bridge and the applied voltage using a digital oscilloscope. Before each run, the volume of liquid is adjusted to have the bridge interface as horizontal as possible, approximating the shape of a cylinder in its state of maximum elongation and electrification. The volumes varied in a range from 20 to 30 microliters. For larger volumes the bridge would bent too soon due to gravitational effects; for smaller ones the bridge would thin too much in the middle, producing unstable necks that might lead to an early pinch off\cite{Ramos:1994}. The voltages are limited in all the experiments done to avoid the build up of high currents, which can lead to several inconveniences and irreversible situations, as will be discussed further on. For this reason, the voltages in this set up could not be risen as much as in the beaker configuration, and therefore the aspect ratios are kept at moderate values.

\begin{figure}
\centering
\includegraphics[width=3.4 in]{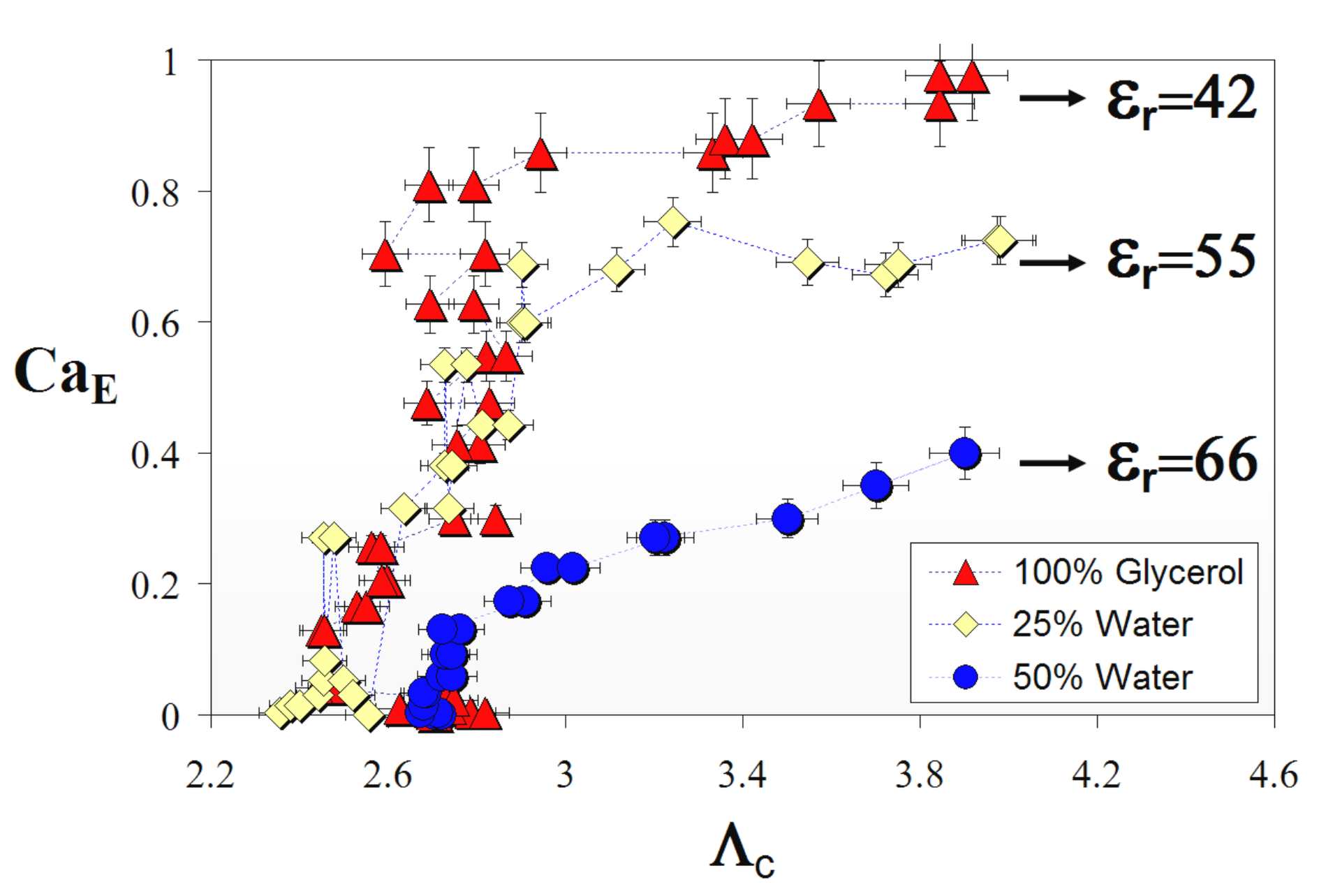}
\caption{ Electrocapillary number $Ca_E$ vs the critical aspect ratio $\Lambda_c=l_c/D_m$ at which the bridge breaks for a fixed voltage and slowly increasing $l$. \label{fig:capillary}}
\end{figure}

With the experimental measurements in this configuration it is possible to compute precisely the Electrocapillary number $Ca_E$ and the critical aspect ratio $\Lambda_c$, which have been plotted in figure \ref{fig:capillary}. Each measurement is repeated at least 3 times, and its dispersion constitutes the error bars shown. Note that unlike the rest of the measurements in the paper, these ones involve critical points where a transition occurs. They are therefore very sensitive to external perturbations and although extreme care had been taken, fluctuations were unavoidable. However, the qualitative result is consistent and reproducible in different runs over different days: it can be observed how the values needed for the Electrocapillary number to create a slender bridge for $\epsilon_r=65.4$ are less than half of those required for pure glycerin $\epsilon_r=41.6$. The behavior is therefore qualitatively analogous to that found for pure dielectric liquids \cite{Gonzalez1989}\cite{Saville:2002}. The same experimental procedure was followed but adding now small concentrations of salt in pure glycerin. This leads to a decrease in stability as the resistivity decreased, as was already observed in the beaker configuration \cite{japos:2009} and predicted before in the literature \cite{Saville:2002}\cite{Saville:1971}. With electrical conductivities of the order as those found in regular tap water (above $\sim 100 \mu S/cm$), the bridges could not be reproduced anymore. Additionally, as observed and studied by Ramos et al.\cite{Ramos:1994}\cite{Gonzalez1989}, the bridge is always observed to become asymmetrical in the instants before breakup, with a more prominent bulb on one side than the other.

Some drawbacks need to be mentioned concerning the experimental results: In an ideal case without gravity and without wetting effects, the liquid cylinder at zero voltage should have a maximum aspect ratio of value $\pi$, according to the Plateau criterium \cite{Plateau}. However, due to gravity and wetting effects, the aspect ratios found in the absence of electric voltage are much lower than those ideally expected \cite{Ramos:1994} \cite{Saville:2002}, namely around 2.4 rather than $\pi$. Secondly, as was already observed in the beaker configuration, hydrogen bubbles are generated at the cathode surface when water-based liquids are used, evidencing the presence of electrolysis in the process, specially when higher currents are generated. This is not an issue in the beaker configuration, where larger volumes of liquid are employed and more surface is exposed to air, hence the bubbles rise to the liquid surface and only rarely they enter into the bridge. In the case of the axisymmetric configuration, covering the electrodes with platinum foil reduced the problem, but did not solve it completely, and bubbles were still generated as the percentage of water and the voltages increased. For this reason, the maximum percentage of water used in the experiments was $50\%$ and the voltages were limited to avoid high currents. Joule heating can be safely neglected in these cases, since the electrical conductivity is reduced significantly as well as the thermal conductivity of the water/glicerine mixtures as compared with the experiments in section \label{section:beakers}. Another issue to have in mind involves the relative importance of gravitational forces in this whole process. Although it has been shown that the Bond numbers are small in these stages, they can not be completely neglected. Several efforts were done in the past to avoid this effect by using density matching techniques\cite{Gonzalez1989} or parabolic flights to perform the experiments in microgravity conditions \cite{Saville:2000}. Finally, the last issue was already mentioned by Melcher and Taylor in the abstract of their pioneering review \cite{MelcherTaylor:1969}, and it concerns the control in polar liquids of the electrical conductivity. Polar liquids are more prone to become contaminated, even in this set up in which they were confined avoiding the contact with contaminants. For this reason, glycerin was chosen as base liquid, and permitted us to achieve more reproducible results. It has been necessary to employ the axisymmetric configuration and different liquids to isolate the dielectric effects from the rest, although the ``floating water bridge'' effect is not as impressive as it was in the beaker configuration due to the mentioned reasons.

Nonetheless, within the aforementioned limitations, we can conclude that induced dielectric polarization is responsible for the stability against capillary collapse in the water bridge through the mechanism described at the beginning of this section. There are still several matters to be clarified: first of all, the role of charge relaxation effects on the instability of the bridge. It has been demonstrated that the increase of free charge (electrical conductivity) certainly disturbs the equilibrium, as was also shown by Burcham \& Saville \cite{Saville:2002} in bridges of dielectric liquids. But, how disturbing can this be? Taking water as a purely dielectric liquid (no free charge) of $\epsilon_r=81$ and performing a simple stability analysis as that performed by Nayyar \& Murty \cite{Nayyar1960}, with $Ca_E \approx 1$, the analysis yields aspect ratios of $\Lambda \approx 100$ (taking for the maximum length of the bridge the maximum unstable wavelength). In contrast, the highest aspect ratios observed experimentally rarely reach values of 10 (right images in figure \ref{fig:beakersmontage}). According to this line of reasoning a non trivial question arises: why is the effect not observed with dielectric bridges in which the free charge can be almost neglected (e.g. oils)? Recurring again to the stability analysis for a dielectric liquid of $\epsilon_r=2$ and a diameter of $2mm$ held in air, even for the maximum electrical fields that can be achieved in our system before air breakdown (some kV per mm), only small aspect ratios are obtained, hardly 5\% higher than the values expected in the absence of electrical field \cite{Saville:1997}. This fact manifests the high dependance of the phenomenon on the dielectric permittivity of the liquid, which is unavoidably connected with the amount of free charge that can be dissolved in the liquid.

The effect of the electrical shear stresses on the stability of the bridge deserves also some comments. It is well known that a strong shear over an interface can be used to stabilize liquid jets \cite{Tomotika}\cite{Saville:1970}\cite{Taylor:1969} \cite{MelcherWarren:1971}. In these cited cases the electrical current is driven by convection and therefore is independent of the downstream conditions. Charge is well separated before reaching the jet\cite{Mora:2007}, and is forced to be transported only in the direction of the flow, under strong surface shear stresses and high accelerations, achieving supercritical regimes\cite{MelcherWarren:1971}. However, for the case of liquid bridges the charge transport is only due to conduction, charge of opposite signs coexist in the system and the electrical shear stresses are nonuniformly and unsteadily distributed along the bridge, giving rise to the complicated flow patterns. In this sense, the conditions and the characteristics of the liquid bridge would resemble those of the ``decelerating stream'', depicted and analyzed by Melcher \& Warren \cite{MelcherWarren:1971}.

\section{Axisymmetric configuration: stability against gravity.}\label{section:gravity}

\begin{figure*}
\includegraphics[width=6.8 in]{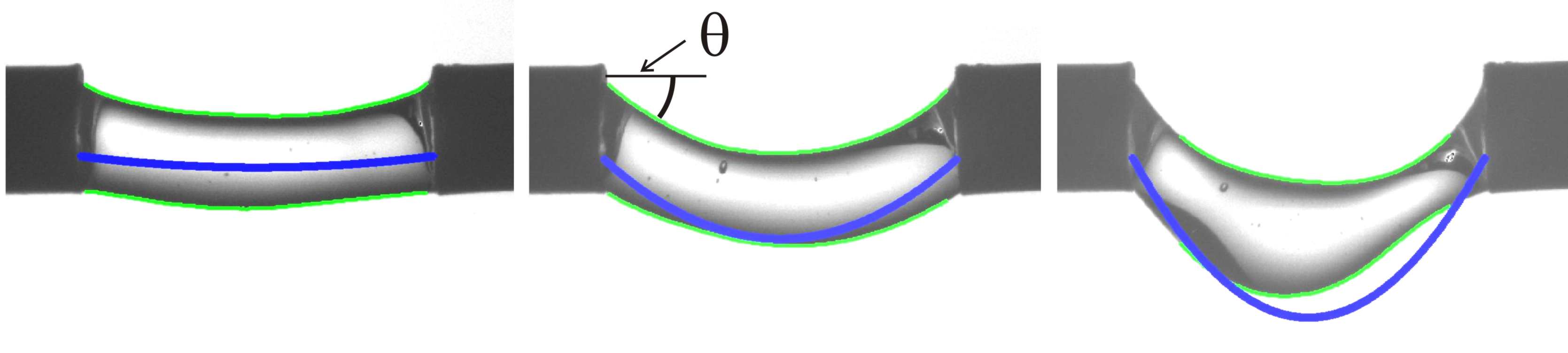}
\caption{ Different side views of a glycerine bridge for decreasing voltages (increasing Electrogravitational numbers $G_E$), from left to right: $G_E= 0.2, 0.8$ and  $1$. A line outlines the liquid surface and the line departing from the middle of the nozzle depicts a parabola with the predicted $\theta$ at the nozzle. The nozzles diameter is 3 mm.\label{fig:gravitymontage}}
\end{figure*}

The previous experiments dealt with the stability of the liquid bridge against capillary forces. In this section a simple experiment will be discussed to study the stability against gravity. The experiment is being carried out in the same setup, but the procedure is changed in this case. For a fixed aspect ratio $\Lambda$, we start applying a voltage $U$ well above the minimum critical value and reduce it until the bridge collapses due to its own weight, in contrast with the capillary collapse discussed in last section. Some sequences of the process can be observed in figure \ref{fig:gravitymontage}. The visualization of the process is synchronized with the voltage measurements, using image processing, a polynomial curve is fitted to the interfaces of the bridge and is used to calculate the angle at the edges of the bridge. The process is performed slow enough to assume that the bridge is in equilibrium when its angle is measured. From the side view we are able to detect four extremes (two at the interface above and other two at the lower interface) and the experimentally measured $\theta$ is the average of the four of them. As we decrease the voltage (increasing $G_E$), the bridge will bend more and more, with increasing $\theta$. These values are plotted against the Electrogravitational number $G_E$ in figure \ref{fig:gravitygraph}. For analyzing this set of data we employ an argument already used in the literature \cite{Widom:2009}, but here it will be adapted to the present situation and further developed. The balance of normal stresses in the liquid bridge interface can be written in a simple form as \cite{CastellanosEHD}:

\begin{equation}
P_i-P_o + \frac{1}{2}(\epsilon_i -\epsilon_o) E_t^2-\frac{1}{2} (\epsilon_i {E^i_n}^2-\epsilon_o {E^o_n}^2)=\frac{\gamma}{R} \label{eq:normal}
\end{equation}
Here $P_i-P_o$ is the pressure jump across the interface, and the indexes $i$ and $o$ stand for \textit{inner} and \textit{outer} respectively, $\epsilon=\epsilon_o \epsilon_r $ is the dielectric permittivity of each medium, $E_n$ and $E_t$ are respectively the normal and tangential components of the electrical field at the interface. In the simplistic case of a purely dielectric liquid, no surface charge exists at the interface. The external electrical field is applied parallel to the interface, and no normal components of the electrical field are induced ($E_n \approx 0$). Under these conditions, the liquid bridge can be here taken as a stable and steady ``viscous catenary'' consisting of a flexible line of mass $\rho V$, subjected to a tension $T$, where the classical force balance

\begin{equation}
\rho V g=2Tsin\theta
\label{eq:catenary}
\end{equation}

must be satisfied in every point, $\theta$ referring to the angle formed with the horizontal, being maximum at the extremes of the line of mass. In the following we will make the non-trivial assumption that the tension in the catenary can be formulated as the overpressure inside the liquid bridge, its nature being mainly electrical, since we will mainly work with Electrocapillary numbers greater than unity and the capillary term in equation \ref{eq:normal} can be neglected in this section. Such a rough assumption was already employed by Widom et al. \cite{Widom:2009} in a different way. In the following it will be carefully compared with the experimental results. Introducing the electrical term in equation \ref{eq:catenary} and rearranging the terms we end up with the following expression:

\begin{figure}
\includegraphics[width=3.4 in]{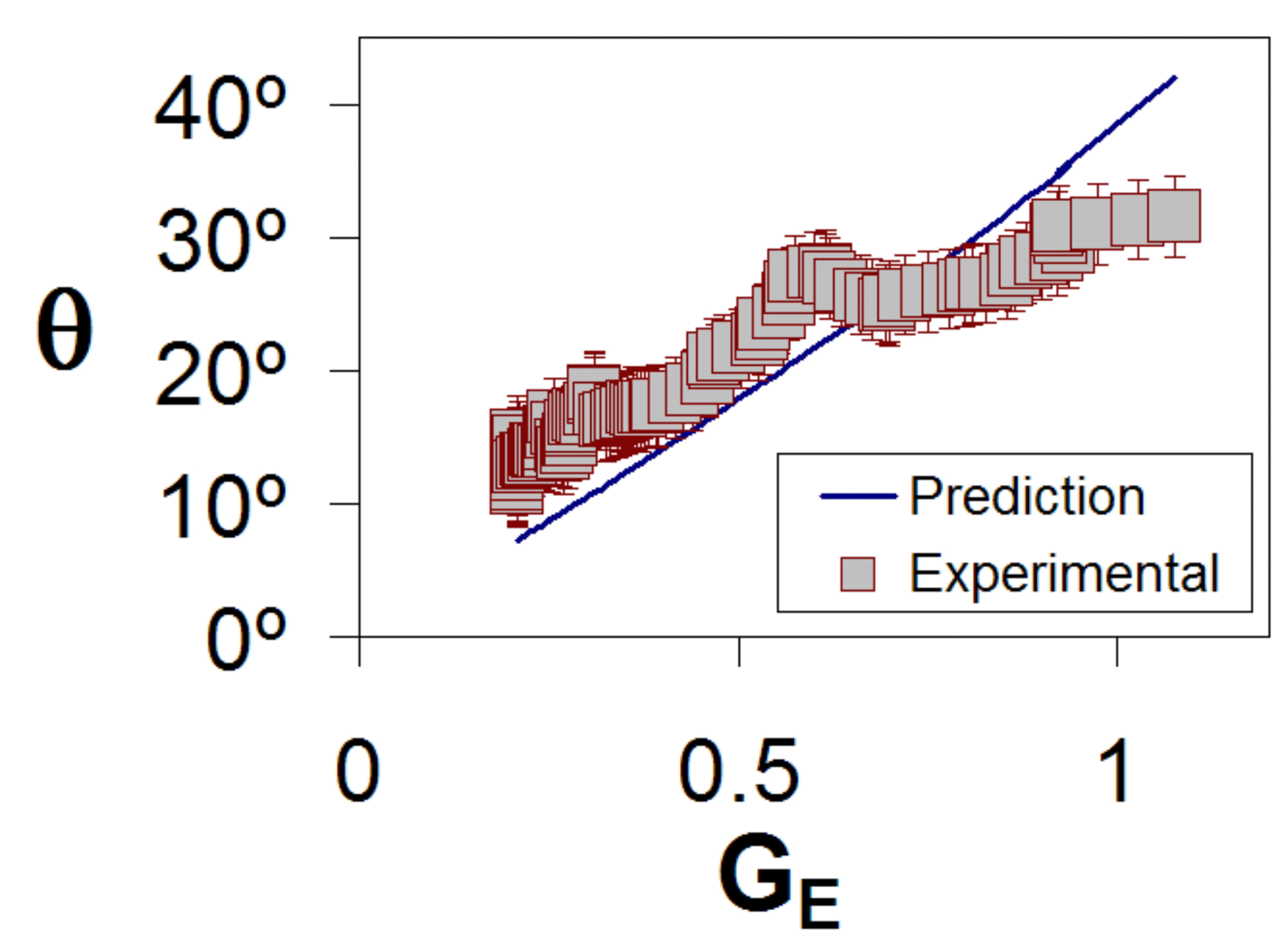}
\caption{ Angle at the edge $\theta$ of the `catenary bridge' vs the Electrogravitational number $G_E$, the straight line corresponds to the prediction in equation \ref{eq:prediction}. \label{fig:gravitygraph}}
\end{figure}

\begin{equation}
sin \theta= \frac{G_E}{2} \left(\frac{\Lambda^{2}}{2\pi}\right)^{1/3}\label{eq:prediction}
\end{equation}
The comparison of this prediction with the experimental data is shown in figure \ref{fig:gravitygraph}, where the experimentally measured angle is compared with the predicted ones in equation \ref{eq:catenary}for different Electrogravitational numbers. In figure \ref{fig:gravitymontage} a parabola departing from the edges with the predicted $\theta$ in equation \ref{eq:catenary} is plotted for comparison with the observed bridge line. A good agreement is found within the experimental errors, which are defined as the dispersion of angle values found between the four bridge edges visible from the side view. However, the values of the angles are slightly underestimated for highly electrified cases, i.e. the role of the electrical forces is overestimated, presumably because the dielectric assumption breaks down. Considering that free charges can be partially stabilized at the surface, the normal components of the electrical field would enter into play in equation \ref{eq:prediction}, probably screening the external longitudinal field. In consequence, it is observed in the experiments that the agreement becomes worse as we increase the conductivity of the liquid, either by adding water or salt.

\section{Flow in the floating water bridge}\label{section:flow}

In the analysis performed so far, the problem has been treated as hydrostatic. However, there is indeed flow of liquid within the bridge in both directions. No characteristic or reproducible flow patterns could be observed. Only in the case of the beaker configuration, when the voltage is set one can see preferred flow directions, leading to readjustments of the water level on each beaker.

As we know from Electrohydrodynamics, an electrical field can not induce a flow at the interface of a purely conducting liquid or a purely dielectric one, since no interfacial shear stress can exist in these ideal cases. The situation changes with a more realistic liquid in which surface charge is not fully equilibrated at the surface, i.e. charge relaxation time is finite, and shear stresses develop over the surface. These interfacial electrical shear stresses are able to induce high velocities, as in the case of the Taylor pump\cite{MelcherTaylor:1969}, electrically driven jets\cite{Taylor:1969}\cite{MelcherWarren:1971}, electrospinning \cite{HohmanBrenner:2001}, and can even induce highly strained flows able to generate geometrical singularities in droplets \cite{Alvi:pof}. Such shear stress depends linearly on the tangential field at the interface and on the surface charge induced at the interface. Here we face a common problem in Electrohydrodynamics: Only in very few paradigmatic experiments the surface charge can be properly modeled and indirectly determined, as in the case of the Taylor pump \cite{MelcherTaylor:1969}\cite{Basaran:1988}. In the particular case of a liquid bridge, Burcham and Saville \cite{Saville:2002} numerically solved the problem for a leaky dielectric, introducing the surface conductivity as a free parameter and they found an heterogenous distribution of charge at the bridge surface, which in their case gave rise to a recirculating flow pattern.


In our case, the distribution of the surface charge and its stability is quite unclear, but according to the conclusions of the last sections, it seems that surface charge is not fully in equilibrium at the surface. Therefore, the electrical shear stress generated must be quite unsteady and inhomogeneous, giving rise to strong spatial and temporal variations in the velocity fields, even giving the impression to be chaotic. However, the temporal scale in which such flow patterns change can be of the order of seconds. This fact permitted to measure the interfacial velocity by using high speed cameras and perform Particle Image Velocimetry in an area close to the surface, combined with Particle Tracking Velocimetry in those cases where the particle density was too reduced.

In order to give a rough prediction on the velocities in the bridge we need an estimate of the electrical shear stress generated at the surface. The balance of shear stresses at the interface can be expressed as follows:

\begin{equation}
q_s E_t=\mu \frac{\partial u_s}{\partial y} \label{eq:shear}
\end{equation}

where $q_s$ refers to the surface charge density, $E_t$ to the tangential field at the surface, $\mu$ to the liquid viscosity and $u_s$ the surface velocity. The surface charge density $q_s=\epsilon_o E^o_n-\epsilon_i E^i_n$ would acquire its maximum value for the perfectly conducting case, in such a case $q_s^o = \epsilon_o E^o_n$ and the surface charge would remain in perfect equilibrium. Unfortunately, nothing can be said a priori about surface charge but $q_s<<q_s^o$, as has been argued in previous sections. Therefore, $u_s$ and $q_s$ can not be estimated independently. To get a better insight into this matter, the following magnitudes ratios are defined:

\begin{align}
\Phi&=\frac{E^i_n}{E^o_n}\\
\Xi&=\frac{E^o_n}{E_t}\label{eq:efield}
\end{align}

With this definition $\Xi$ express the relative importance of the normal electrical field components against the tangential ones, responsable for the shear stress. And $\Phi$ goes to zero for perfectly conducting liquids, in which charge is stable at the interface and no shear stress can exists. Moreover, surface velocity can be expressed in dimensionless units making use of the capillary velocity $u_o=\gamma/\mu$ and the Electrocapillary number $Ca_E$. With these introduced definitions, the induced velocities at the interface can be written as:
\begin{figure}

\includegraphics[width=3.4 in]{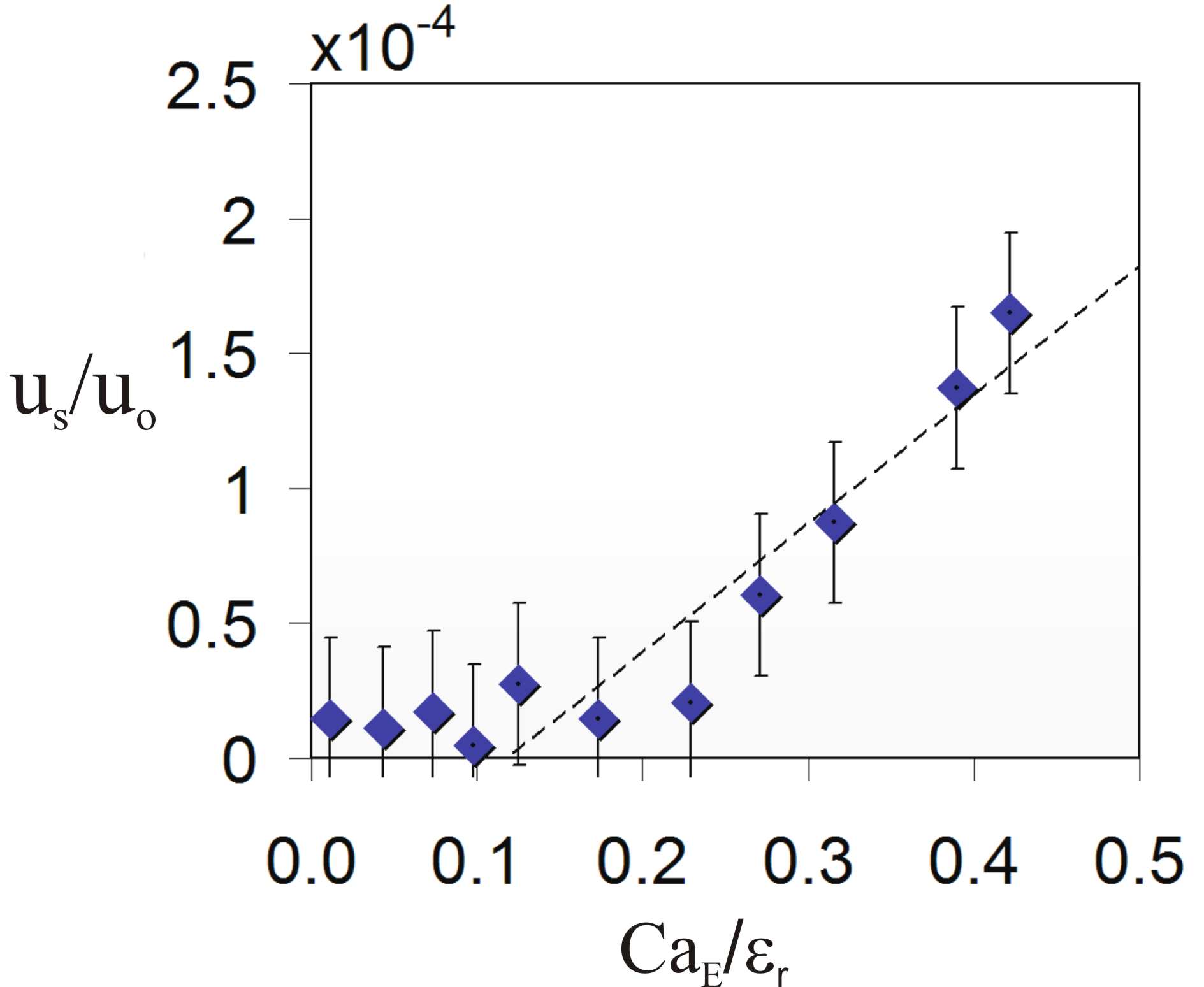}

\caption{Surface velocity for different Electrocapillary numbers $Ca_E$ (blue dots), straight line represent a linear fit.\label{fig:flowgraph}}
\end{figure}

\begin{equation}
\frac{u_s}{u_o} \sim \frac{Ca_E}{\epsilon_r}(1-\epsilon_r\Phi)\Xi\label{eq:velocity2}
\end{equation}

Surface velocities were measured for different applied voltages and plotted in dimensionless units in figure \ref{fig:flowgraph}. The liquid employed was a glycerin-water mixture $80\%-20\%$ ($\epsilon_r\simeq 52$, Electrical Conductivity $0.125\mu S/cm$), seeded with neutrally buoyant polystyrene particles of 20 microns in diameter. ${q_s}/{q_s^o}$ and ${E^o_n}/{E_t}$ can not be independently obtained, but an approximated value can be given to its product through a linear fit from graph \ref{fig:flowgraph}:

\begin{equation}
(1-\epsilon_r\Phi)\Xi=\frac{q_s}{q_s^o}\frac{E^o_n}{E_t}\approx 4.8\times10^{-4}
\end{equation}

As expected, values of surface charge are far from those of a purely conducting liquid, whose charge would be in perfect static equilibrium. But still, the small amount of free charge available at the surface is enough to set the liquid into motion and therefore it can not be fully neglected. This fact represents the main idea behind the ``Leaky Dielectric Model'' of Taylor and Melcher\cite{MelcherTaylor:1969}, which applies perfectly here. Much higher velocities and surface charge can be expected therefore as the amount of water is slightly increased. In addition, the fact that $u_s \ll u_o$ justifies the assumption that the flow does not play a relevant role in the stability of the liquid bridge, and therefore our quasistatic assumption is justified. Regarding the stability of the surface charge, these measurements also confirm that convection times in the system $t_c=l/u_s\sim 1s$ are much longer than charge relaxation times $t_c=\epsilon_r\epsilon_c/K \sim 35ms$; therefore, charge convection can not be able to destabilize surface charge. In conclusion, only surface charge conduction could be responsible for surface charge instability and for the observed disordered flow patterns.

\begin{figure*}
\centering

\includegraphics[width=6.8 in]{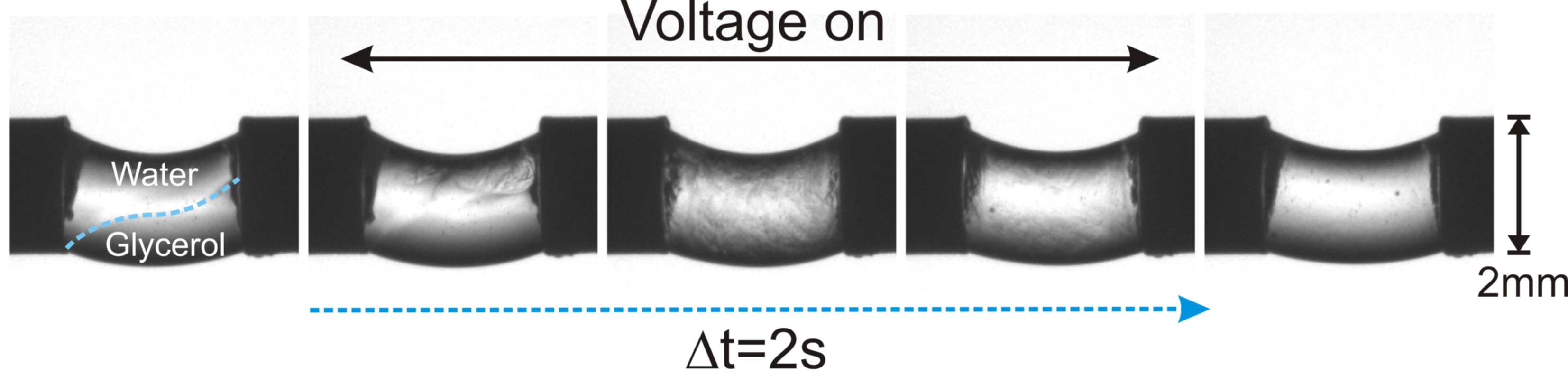}
\caption{ Visualization of the mixing process with $5\mu l$ of water and $5\mu l$ of glycerol (1000 times more viscous). First and last snapshots are the initial and the final state. The total process takes about 2 seconds\label{fig:flowmontage}}

\end{figure*}

Interestingly enough, this result shows us that high velocities of order of several millimeters per second or higher could be reached in bridges of water-based liquids with small volumes of a few microliters. Such a feature could have great benefits for efficiently mixing of small volumes of liquids. In the following experiment, a 5 microliter droplet of deionized water is placed at the exit of one cylinder and 5 microliter droplet of pure glycerin (1000 times more viscous than water) in the opposite, when both establish contact, the glycerine phase stays at the bottom and water above (left image in figure \ref{fig:flowmontage}). Although both liquids are miscible, the difference in viscosities and densities makes the mixing of both liquids a complicated task, impossible to accomplish by diffusion in practical time scales. When the voltage is elevated to 15kV ($Ca_E \approx 10$) the liquids are set into motion following irregular and unsteady flow patterns. As a consequence, the liquids are perfectly mixed within a few seconds. For this application the liquids do not need to be directly connected to the electrodes, and the electrical current through the bridge therefore can be controlled.

\section{Summary}\label{section:summary}

In conclusion, the floating water bridge remains stable without breaking for big aspect ratios and unbent by gravity due to the effect of the induced polarization forces at the interface, generating normal stresses that counteract not only capillary forces but also gravity. To our knowledge, such a balance between polarization effects and gravity has never been reported in the literature with the only exception of the Pellat effect\cite{Pellat}. Free surface charge is responsible for the observed flow, which can be especially intense for aqueous solutions, due to their lower viscosity and their non negligible conductivity. The main difference with previous studies in the field, as those by Gonz\'alez et al.\cite{Gonzalez1989}\cite{Ramos:1994} and Burcham \& Saville\cite{Saville:2002}, consists in the presence of a significant free charge in the system, which remains out of equilibrium.

In spite of the effort of this contribution, the water bridge deserves further investigation. Improved and more sophisticated experimental configurations should be designed to allow for better flow manipulation and measurements.

\begin{acknowledgements}\label{section:thanx}
The authors thank Dr. Elmar Fuchs for making them aware of the phenomenon of the floating water bridge. \'Alvaro G. Mar\'in thanks especially Dr. Pablo Garc\'ia-S\'anchez for long discussions on the subject and also Prof. Antonio Ramos for his useful comments. \'Alvaro G. Mar\'in dedicates this paper to the memory of Prof. Antonio Barrero Ripoll.
\end{acknowledgements}

\end{document}